\documentclass[12pt]{iopart}

\usepackage{braket}
\usepackage{graphicx}
\usepackage{braket}
\usepackage[lofdepth,lotdepth]{subfig}
\usepackage[numbers,square,sort&compress]{natbib}

\usepackage{hyperref}
\hypersetup{dvips,
  pdfauthor={Christopher N. Varney, Kai Sun, Victor Galitski, and
    Marcos Rigol},
  pdftitle={Quantum Phases of Hard-Core Bosons in a Frustrated
    Honeycomb Lattice}, 
  colorlinks=true,
  linkcolor=blue,
  citecolor=blue,
  pdfpagemode=UseNone
}

\newcommand{\bbraket}[1]{\braket{\hspace{-2pt}\braket{#1}\hspace{-2pt}}}

\begin{document}
\title{Quantum Phases of Hard-Core Bosons in a Frustrated Honeycomb Lattice}
\author{C.~N.~Varney$^{1,7}$, K.~Sun$^2$, V.~Galitski$^{3,4}$, and M.~Rigol$^{5,6}$}
\address{$^1$Department of Physics, University of Massachusetts, Amherst,
  Massachusetts 01003, USA}
\address{$^2$Department of Physics, University of Michigan, Ann Arbor,
  MI 48109, USA} 
\address{$^3$Joint Quantum Institute and Department of Physics, University
  of Maryland, College Park, Maryland 20742, USA }
\address{$^4$Condensed Matter Theory Center, Department of Physics,
  University of Maryland, College Park, Maryland 20742, USA}
\address{$^5$Department of Physics, Georgetown University, Washington, DC
  20057, USA}
\address{$^6$Physics Department, The Pennsylvania State University, 104
  Davey Laboratory, University Park, Pennsylvania 16802, USA}
\address{$^7$Corresponding author}
\ead{varney@physics.umass.edu}

\begin{abstract}
  Using exact diagonalization calculations, we investigate the
  ground-state phase diagram of the hard-core Bose-Hubbard-Haldane
  model on the honeycomb lattice. This allows us to probe the
  stability of the Bose-metal phase proposed in Varney {\it et al.},
  Phys. Rev. Lett. {\bf 107}, 077201 (2011), against various changes
  in the originally studied Hamiltonian.
\end{abstract}

\section{Introduction}
\label{sec:intro}
In nature we are surrounded with examples of ordered phases at low
temperatures---e.g.~crystalline solid structures, magnetically ordered
materials, superfluid and superconducting states, etc. While it is
straightforward to think of these ordered phases melting as the
temperature is increased into the familiar classical liquid or gaseous
states that are commonplace in every aspect of our lives, it has been
a long-standing question as to whether `quantum melting' at zero
temperature can act similarly to thermal effects and prevent
ordering. For a quantum spin or boson system, the resulting state of
matter is known as a quantum spin liquid \cite{pomeranchuk1941}. The
interest in such a hypothetical spin liquid has remained strong for
decades, most prominently due to the discovery of high temperature
superconductivity \cite{anderson1987,lee2006}.

Of critical importance is whether a two(or higher)-dimensional system
can host a quantum spin liquid. At present, there exists a complete
classification of quantum orders \cite{wen2002}, which divides
hypothetical spin liquids into several distinct classes. Some
theoretical stability arguments have also been presented showing that
there is no fundamental obstacle to the existence of quantum spin
liquids \cite{hermele2004}. Gapped spin liquid phases have been
observed in dimer models \cite{roksar1988,moessner2001,yao2012}, and
also a family of special exactly-solvable toy models were discovered
which can support gapped and gapless spin liquid phases
\cite{kitaev2006}. Although these discoveries clearly demonstrated
that a spin-liquid phase may appear in two (or higher) dimensions, at
least in toy models, whether the same type of exotic phase can appear
in a realistic spin system remains unclear.

Very recently, there has been much numerical \cite{meng2010,syan2011,
  okumura2010,cabra2011,clark2011,mezzacapo2012,kalz2012,yang2012} and
experimental \cite{vachon2011,yan2012,liu2012} evidence to suggest the
existence of gapped spin liquids in models with $SU(2)$ symmetry, but
it is still unclear why these simple models can support such exotic
phases. Of particular note are the numerical discoveries of a gapped
spin liquid in the Heisenberg model on the kagom\'e lattice
\cite{syan2011} and in the Hubbard model on a honeycomb lattice
\cite{meng2010}. The existence of the latter is especially surprising
and remains under debate \cite{sorella2012}. The nature of this phase
has been the subject of many works and it has been argued that
next-nearest-neighbor exchange coupling is the mechanism responsible
for the quantum spin liquid \cite{cabra2011,clark2011,wang2010,
  lu2011}. However, despite a number of numerical investigations into
this $J_1$-$J_2$ model, there are still open debates on whether the
non-magnetic state present in this model is a valence bond solid
\cite{mulder2010, albuquerque2011,reuther2011,oitmaa2011,mosadeq2011}
or a quantum spin liquid \cite{okumura2010,mezzacapo2012,kalz2012}.

Gapless spin liquids, which may have low-lying fermionic spinon
excitations that strongly resemble a Fermi-liquid state, have remained
more elusive. Because of these excitations and because spin-$1/2$
models can be mapped onto hard-core boson models, some of these
gapless spin liquids are often referred to as a Bose metal or Bose
liquid. The hallmark feature of a Bose metal is the presence of a
singularity in momentum space, known as a Bose surface
\cite{motrunich2007,sheng2009,yang2010,dang2011,mishmash2011,
  varney2011}. However, unlike a Fermi liquid, where the Fermi wave
vector depends solely on the density of the fermions, the Bose wave
vector depends on the control parameters of the Hamiltonian and can
vary continuously at fixed particle density.

In this paper, we follow up on the proposition of such a putative Bose
metal phase in a simple hard-core boson ($XY$) model on the honeycomb
lattice \cite{varney2011}, with an analysis of the stability of this
phase against various changes in the Hamiltonian studied
originally. First, we examine the dependence of the Bose wave vector
on a (phase) parameter that makes the model transition between
frustrated and non frustrated regimes. Next, we show that the phase
identified as a Bose metal is stable to breaking of time-reversal
symmetry and is present in the phase diagram of the hard-core
Bose-Hubbard-Haldane (BHH) model, which features (at least) three
phase transitions.

The remainder of this paper is structured as follows. In
section~\ref{sec:modelmeth}, we define the model Hamiltonian, briefly
discuss the Lanczos algorithm, and define the key observables used in
this study: the charge-density wave structure factor, the ground state
fidelity metric, and the condensate fraction. Next, in
section~\ref{sec:xy}, we discuss the identifying characteristics of
the Bose metal phase in the context of the hard-core boson ($XY$)
model and show how the Bose wave vector evolves as the parameters are
varied. In section~\ref{sec:bhh}, we discuss the three phase
transitions that we can identify in the Bose-Hubbard-Haldane model:
BEC-CDW, BEC-Bose metal, and Bose metal(other phase)-CDW. The main
results are summarized in section~\ref{sec:conc}.

\section{Model and Methods}
\label{sec:modelmeth}
\subsection{Models}
\label{sec:model}
The model proposed in \cite{varney2011} to exhibit a Bose metal
phase is the spin-$1/2$ frustrated antiferromagnetic-$XY$ model on the
honeycomb lattice
\begin{equation}
  H = J_1 \sum_{\braket{ij}} ( S_i^+ S_j^- + \textrm{H.c.} ) + J_2
  \sum_{\bbraket{ij}} ( S_i^+ S_j^- + \textrm{H.c.})
  \label{eq:Ham_xy} \,.
\end{equation}
where $S_i^\pm$ is an operator that flips a spin on site $i$ and $J_1$
($J_2$) is the nearest-neighbor (next-nearest-neighbor) spin
exchange. In this model, the next-nearest-neighbor coupling introduces
frustration as long as $J_2 > 0$ (antiferromagnetism).

The Hamiltonian in \eref{eq:Ham_xy} maps to a hard-core boson model
($S_i^+ \to b_i^\dag$, $S_i^- \to b_i^{\phantom\dag}$, and $J_i \to
t_i$),
\begin{equation}
  H = t_1 \sum_{\braket{ij}} ( b_i^\dag b_j^{\phantom\dag} + \textrm{H.c.} ) +
  t_2 \sum_{\bbraket{ij}} ( b_i^\dag b_j^{\phantom\dag} + \textrm{H.c.} ).
\label{eq:Ham_boson}
\end{equation}
Here $b_i^\dag$ ($b_i^{\phantom\dag}$) is an operator that creates
(annihilates) a hard-core boson on site $i$ and $t_1$ ($t_2$) is the
nearest-neighbor (next-nearest-neighbor) hopping amplitude. This
Hamiltonian was shown to feature four phases: a simple Bose-Einstein
condensate (BEC) [a zero momentum (${\bf k} = 0$) condensate], a Bose
metal (a gapless spin liquid), and two fragmented BEC states. The Bose
metal (BM) was found to be the ground state of this model over the
parameter range $0.210(8) \le t_2 / t_1 \le 0.356(9)$
\cite{varney2011}.

To better understand the stability of the latter phase, we consider a
strongly interacting variant of the Haldane model \cite{haldane1988},
the hard-core Bose-Hubbard-Haldane Hamiltonian \cite{varney2010}
\begin{equation}
  H = t_1 \sum_{\braket{ij}} ( b_i^\dag b_j^{\phantom\dag} + \textrm{H.c.} ) +
  t_2 \sum_{\bbraket{ij}} ( b_i^\dag b_j^{\phantom\dag} e^{i \phi_{ij}} +
  \textrm{H.c.} ) + V \sum_{\braket{ij}} n_i n_j,
  \label{eq:Ham_BHH}
\end{equation}
which reduces to \eref{eq:Ham_boson} for $\phi_{ij}=0$ and $V = 0$.
Here, $V$ describes a nearest-neighbor repulsion and the next-nearest
neighbor hopping term has a complex phase $\phi_{ij} = \pm \phi$,
which is positive for particles moving in the counter-clockwise
direction around a honeycomb. Note that the Hamiltonian in
\eref{eq:Ham_BHH} can be mapped to a modified $XXZ$-model ($S_i^+ \to
b_i^\dag$, $S_i^- \to b_i^{\phantom\dag}$, $n_i \to S_i^z+1/2$, $t_1
\to J_1$, $t_2 \to J_2$, and $V \to J_z$)
\begin{equation}
  \fl H = J_1 \sum_{\braket{ij}} ( S_i^+ S_j^- + \textrm{H.c.} ) + J_2
  \sum_{\bbraket{ij}} ( S_i^+ S_j^- e^{i \phi_{ij}} + \textrm{H.c.}) +
  J_z \sum_{\braket{ij}} \left( S_i^z + \frac{1}{2} \right) \left(
    S_j^z + \frac{1}{2} \right) .
  \label{eq:Ham_xxz}
\end{equation}

The complex phase $\phi$ plays two important roles. Firstly, for $\phi
\ne n \pi$, time-reversal symmetry is explicitly broken. Therefore, we
can use this control parameter to study the stability of the Bose
metal phase against time-reversal symmetry breaking. Secondly, in the
spin language, as we increase the value of $\phi$ from $0$ to $\pi$,
the sign for the next-nearest-neighbor spin-spin interaction is
flipped from positive ($\phi=0$) to negative ($\phi=\pi$), i.e., the
next-nearest-neighbor spin exchange changes from antiferromagnetic to
ferromagnetic. Since frustration in this model originates from the
antiferromagnetic next-nearest-neighbor spin exchange, we can use
$\phi$ to tune the system from a frustrated ($\phi=0$) to a
non-frustrated ($\phi=\pi$) regime, and thus it enables us to explore
the role of frustration in stabilizing the BM phase.

In what follows, $t_1 = 1$ sets our unit of energy, and we fix $t_2 =
0.3$ to focus on transitions from the phase identified in
\cite{varney2011} as a BM phase. This model has two limiting cases:
(1) for $V \to \infty$, the Ising regime, the ground state is a charge
density wave (CDW) and (2) for $V = 0$ and $\phi = \pi$, the
non-frustrated regime, the ground state is a simple zero-momentum BEC
with non-zero superfluid density (SF).

\subsection{Method and Measurements}
\label{sec:method}
To determine the properties of the ground state of \eref{eq:Ham_BHH},
we utilize a variant of the Lanczos method \cite{Lanczos1950}.  This
technique provides a simple and unbiased way to determine the exact
ground-state wave-function for interacting Hamiltonians.  One
limitation of the original algorithm is that the Lanczos vectors may
lose orthogonality, resulting in spurious eigenvalues
\cite{Cullum1985}. Orthogonality can be restored through
reorthogonalization \cite{Simon1984}, which requires storing the
Lanczos vectors. The storage needs can then be reduced utilizing a
restarting algorithm, and the most successful techniques are the
implicitly restarted \cite{Sorenson1992, Calvetti1994} and the
thick-restart Lanczos algorithms \cite{KWu2000}. These two methods are
equivalent for Hermitian eigenvalue problems, and here we utilize the
thick-restart method for its simplicity in implementation.

A generic and unbiased way of determining the location of a quantum
phase transition is related to the ground-state fidelity metric, $g$
\cite{varney2010,zanardi2006,campos2008,rigol2009,gu2010}. The
fidelity metric is a dimensionless, intensive quantity and is defined
as
\begin{equation}
  g = \frac{2}{N} \frac{1 - F(\lambda, \delta\lambda)}{(\delta
    \lambda)^2},
\end{equation}
where $N$ is the number of sites and the fidelity
$F(\lambda,\delta\lambda)$ is
\begin{equation}
  F(\lambda, \delta\lambda) = \braket{\Psi_0(\lambda) |
    \Psi_0(\lambda+\delta\lambda)},
\end{equation}
where $\ket{\Psi_0(\lambda)}$ is the ground state of $H(\lambda)$, and
$\lambda$ is the control parameter of the Hamiltonian.

For strong repulsive interactions, the ground state of the BHH model
is a charge-density-wave (CDW) insulator, where one of the two
sublattices is occupied while the other one is empty. This state
spontaneously breaks the six-fold rotational symmetry down to
three-fold but leaves the lattice translational symmetry intact. In
addition, because of the diagonal character of the order established,
the structure factor that describes this phase is maximal at zero
momentum. Thus, we define the CDW structure factor $S_{\rm CDW}$ as
\begin{equation}
  S_{\rm CDW} = \frac{1}{N} \sum_{i,j} \braket{(n_i^a - n_i^b)(n_j^a - n_j^b)},
\end{equation}
where $n_i^a$ and $n_i^b$ are the number operators on sublattice $a$
and $b$ in the $i$th unit cell, respectively.

Another possible ordered state is a Bose-Einstein condensate, where,
in our model, bosons can condense into quantum states in which
different momenta are populated. According to the Penrose-Onsager
criterion \cite{penrose1956}, the condensate fraction can be computed
by diagonalizing the one-particle density matrix $\rho_{ij} =
\braket{b_i^\dag b_j^{\phantom\dag}}$,
\begin{equation}
  f_c = \Lambda_1 / N_b,
\end{equation}
where $\Lambda_1$ is the largest eigenvalue of $\rho_{ij}$ and $N_b$
is the total number of bosons. In a BEC, the condensate occupation
scales with the total number of bosons as the system size is
increased, which is equivalent to stating that $\rho_{ij}$ exhibits
off-diagonal long-range order \cite{yang1962}. Consequently, in a
simple BEC, $\Lambda_1 \sim O(N_b)$ while all other eigenvalues are
$O(1)$ \cite{leggett2001}. Aside from a simple BEC, the eigenspectrum
of the single-particle density matrix can signal fragmentation, where
condensation occurs to more than one effective one-particle state
\cite{leggett2001,stanescu2008}, and the Bose metal phase. In the
former case, some of the largest eigenvalues are $O(N_b)$ and could
even be degenerate. For the Bose metal, however, all of the
eigenvalues of $\rho_{ij}$ are $\sim O(1)$. Thus finite size scaling
of $f_c$ can help pinpoint the presence or absence of condensation.

\begin{figure}[t]
  \centering
  \includegraphics[height=0.600\linewidth,angle=-90]{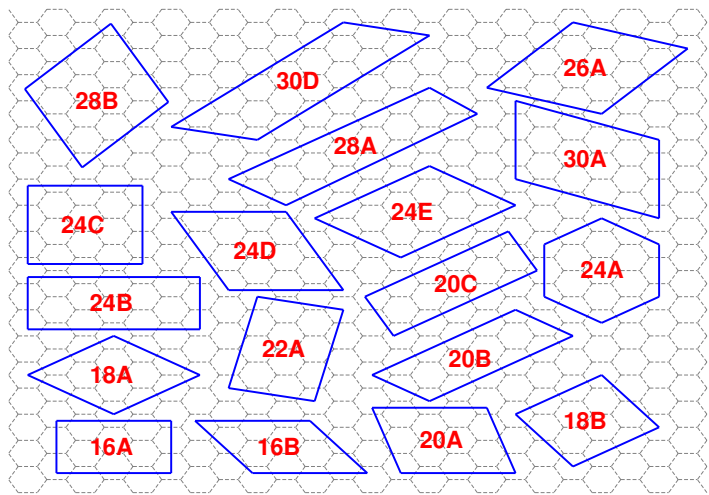}
  \caption{
    Clusters used in this study.
    \label{fig:clusters}
  }
\end{figure}

Further understanding of the latter two phases can be gained by
calculating the single-particle occupation at different momentum
points
\begin{equation}
  n({\bf k}) = \braket{\alpha_{\bf k}^\dag \alpha_{\bf
      k}^{\phantom\dag}} + \braket{\beta_{\bf k}^\dag \beta_{\bf
      k}^{\phantom\dag}},
\end{equation}
where $\alpha_{\bf k}^{\phantom\dag} = \sum_{i\in A} e^{i {\bf k}
  \cdot {\bf r}_i} b_i^\dag b_i^{\phantom\dag}$ and $\beta_{\bf
  k}^{\phantom\dag} = \sum_{i\in B} e^{i {\bf k} \cdot {\bf r}_i}
b_i^\dag b_i^{\phantom\dag}$ are boson annihilation operators at
momentum ${\bf k}$ for the $A$ and $B$ sublattices, respectively. In
order to minimize finite-size effects and fully probe the Brillouin
zone we average over $40\times40$ twisted boundary
conditions\cite{poilblanc1991,claudius1992},
\begin{equation}
  \braket{n({\bf k})}_{\theta_x,\theta_y} = \oint d\theta_x \oint d\theta_y
  \braket{n({\bf k},\theta_x,\theta_y)},
\end{equation}
where $\theta_\alpha$ is the flux associated with the twisted boundary
condition.

For any finite-size calculation, there are a large number of clusters
that one could study, each with slightly different symmetry
properties. In this work, we focus solely on clusters that can be
described by a parallelogram or ``tilted rectangle''. The clusters
used in this study are illustrated in figure~\ref{fig:clusters} and are
discussed in more detail in \cite{varney2010} and
\cite{varney2011}. 

\section{$XY$ Model}
\label{sec:xy}
In a previous study \cite{varney2011}, we reported that the phase
diagram of the $XY$ model~\eref{eq:Ham_xy} on a honeycomb lattice
has three quantum phase transitions separating four distinct
phases. The four phases are: (i) a BEC ${\bf k}=\Gamma$
(antiferromagnetism), (ii) a Bose metal (spin liquid), (iii) a BEC at
${\bf k}=M$ (a collinear spin wave), and (iv) a BEC at ${\bf k}=K$
($120^\circ$ order).

\begin{figure}[tb]
  \centering$
  \begin{array}{ccc}
    \includegraphics[height=0.175\textheight,draft=false]{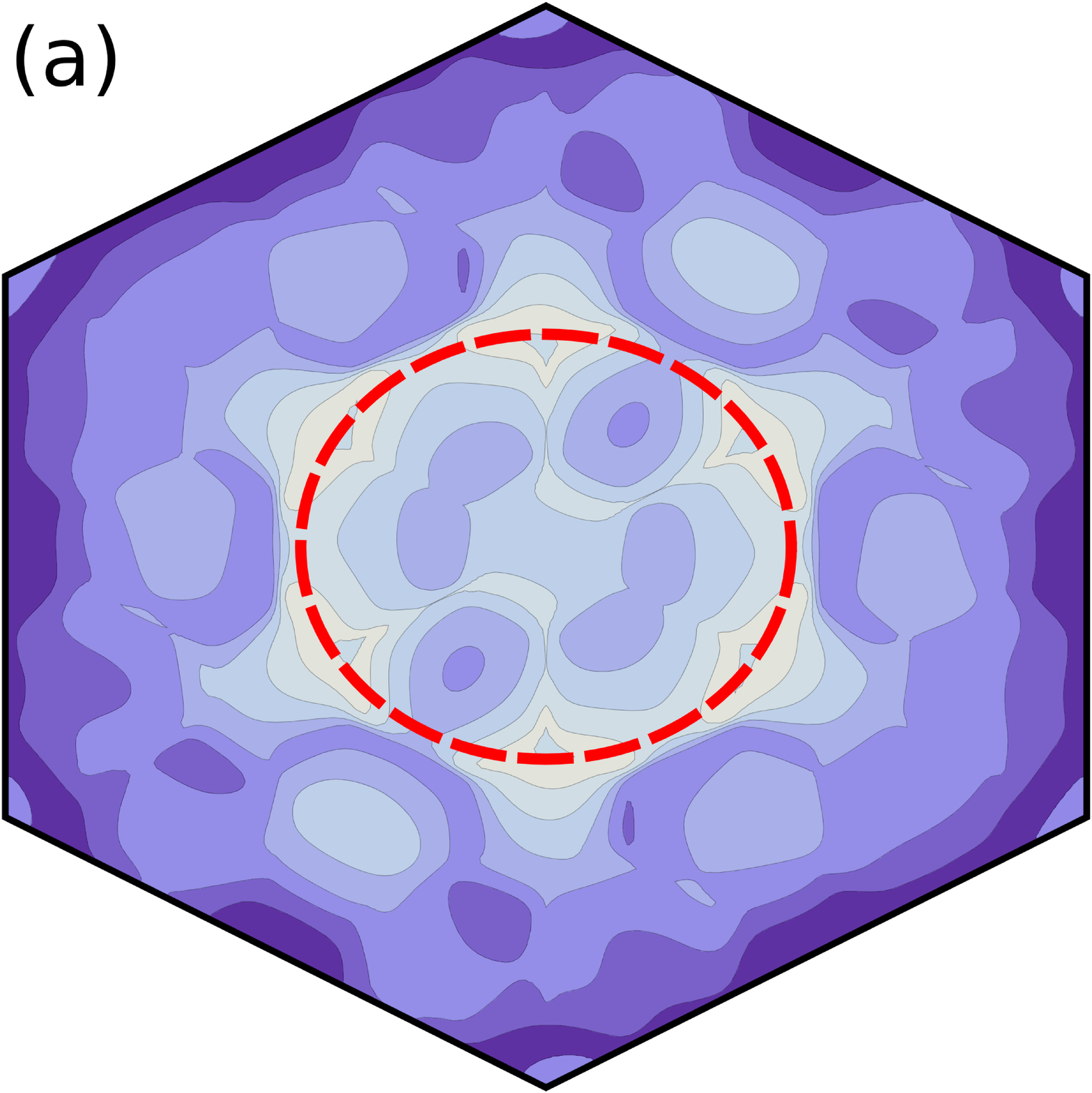} &
    \includegraphics[height=0.175\textheight,draft=false]{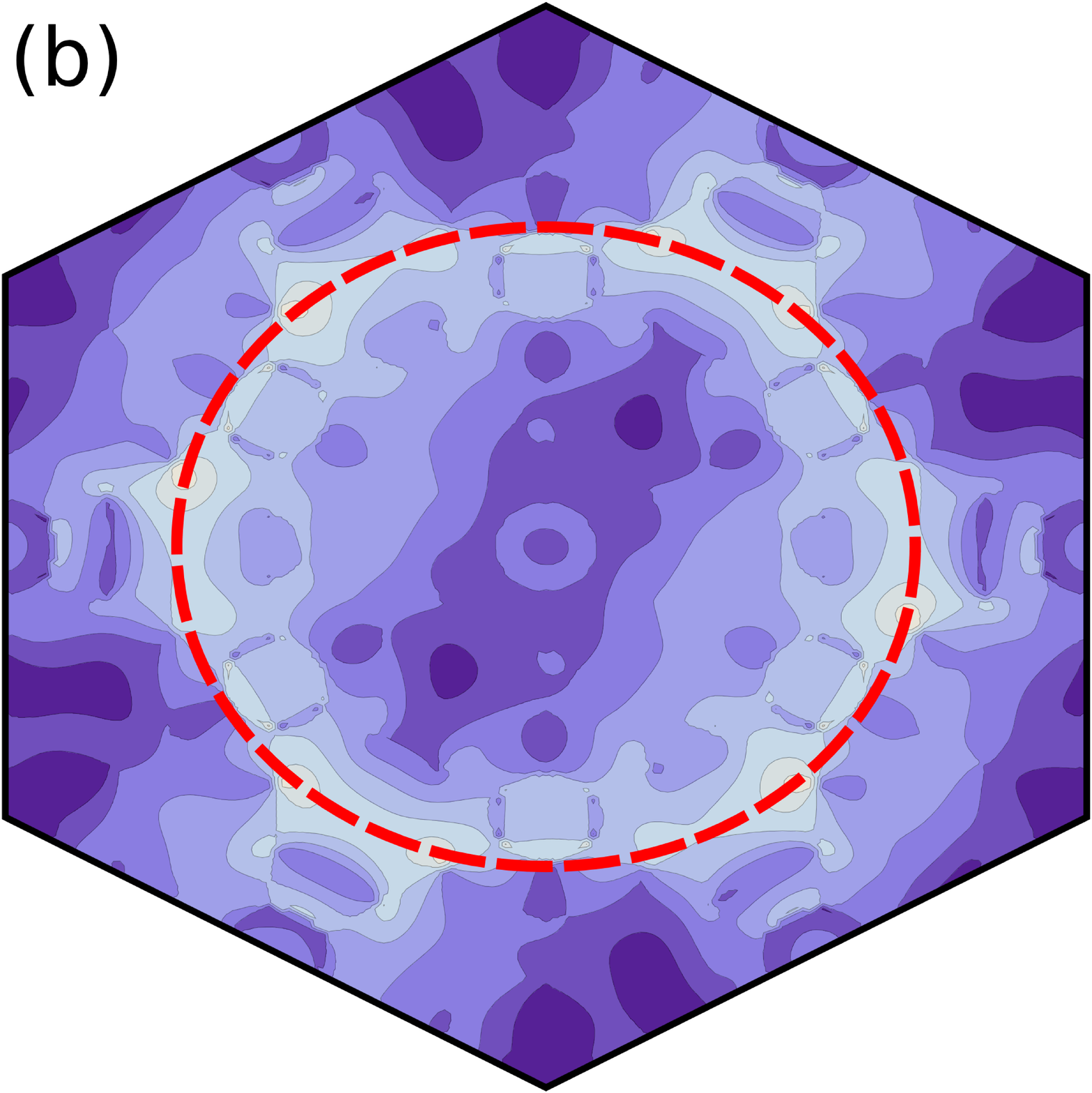} &
    \includegraphics*[height=0.175\textheight,draft=false,viewport=5 11 354 242,draft=false]{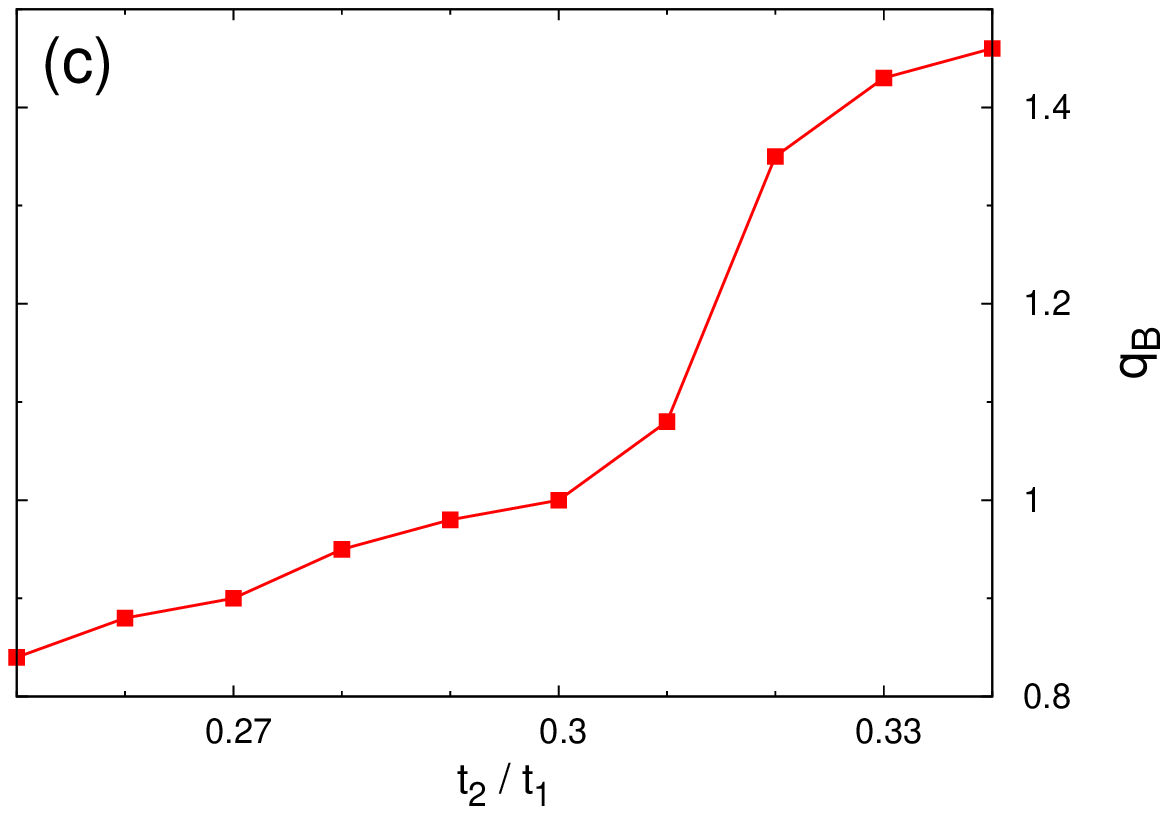}
  \end{array}$
  \caption{
    (Color online) (a)-(b) Momentum distribution $n({\bf k})$ versus
    ${\bf k}$~\cite{footnote2} for the Bose metal phase in the $XY$
    model for $t_2 / t_1 = 0.28$ and $0.33$, respectively. In both
    panels, $40\times40$ twisted boundary conditions were averaged to
    generate $n({\bf k})$, and the Bose surface is indicated by a
    dashed red line. (c) The  magnitude of the Bose-surface $q_B$ as a
    function of $t_2 / t_1$. 
    \label{fig:bosesurface}
  }
\end{figure}

The key signature of a Bose metal is the absence of any order and a
singularity in the momentum distribution $n({\bf k})$. In
figure~\ref{fig:bosesurface}(a) and \ref{fig:bosesurface}(b), we show
$n({\bf k})$ for two values of $t_2/t_1$ that are typical for the Bose
metal phase. For this phase, $n({\bf k})$ features a
$t_2/t_1$-dependent Bose surface, which, as a guide to the eye, we
indicate by a dashed red line. In general, the Bose wave vector $q_B$
at which the maxima of $n({\bf k})$ occurs increases with increasing
$t_2/t_1$, as shown in figure~\ref{fig:bosesurface}(c). We emphasize
that the maxima in $n({\bf k})$ do not reflect Bose-Einstein
condensation as they do not scale with system size.

\section{BHH Model}
\label{sec:bhh}
In this work, we present evidence that, in the $(\phi,V)$ plane [see
\eref{eq:Ham_BHH}], the Bose-Hubbard-Haldane model for $t_2/t_1=0.3$
exhibits (at least) three phases at half-filling. For strong coupling
$V$, the ground state is a charge-density-wave (CDW), while (at least)
two possible ground states exist at weak-coupling. In the frustrated
regime ($\phi \sim 0$) at $V = 0$, the system favors a Bose-metal,
while the unfrustrated regime ($\phi \sim \pi$) favors a
BEC. Consequently, we find that there are three types of transitions:
(i) BEC-CDW, (ii) BEC-BM, and (iii) BM(other phase)-CDW.

\begin{figure}[t]
  \centering
  \includegraphics[width=0.900\linewidth,draft=false]{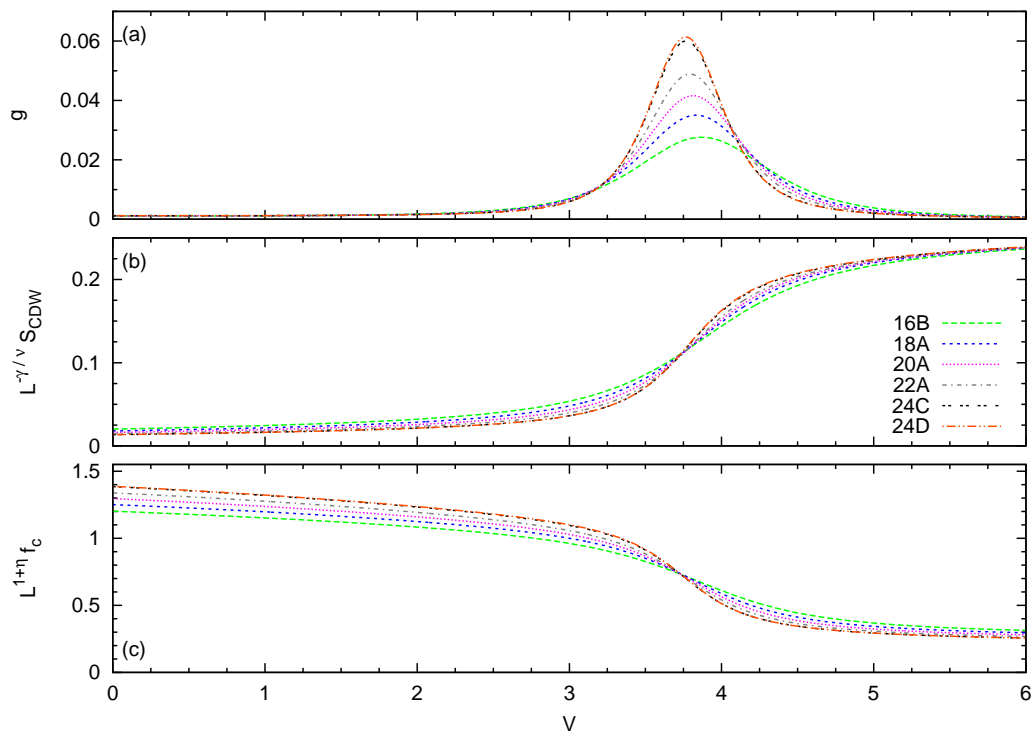}
  \caption{
    (Color online) (a) Fidelity metric $g$, (b) scaled structure
    factor $L^{-\gamma / \nu} S_{\rm CDW}$, and (c) scaled condensate fraction
    $L^{1+\eta} f_c$ as a function of interaction strength for various
    clusters with $\phi = \pi$.
    \label{fig:sfcdw}
  }
\end{figure}

Let us first consider the BEC-CDW transition driven by $V$ at constant 
$\phi$. In figure~\ref{fig:sfcdw}, we show the properties of the system 
for $\phi = \pi$. In panel (a), we show the fidelity metric versus $V$, which 
has a smooth peak that grows with system size, indicative of a second-order 
phase transition (which would be unconventional in this case
in which the system transitions between two ordered states) 
or a weakly first order transition. If the former is true, the structure 
factor would scale according to the rule:
\begin{equation}
  L^{-\gamma / \nu} S_{\rm CDW} = f[(V-V_c) L^{1/\nu}],
\end{equation}
where $N$ is the number of sites, $L=N^{1/2}$ is the linear dimension,
and $\gamma=\nu(2-\eta)$. Because of our small lattice sizes, we
cannot pinpoint the exact nature of this transition. For example,
using a scaling analysis based on the 3D Ising \cite{hasenbusch2010}
and $XY$ universality classes \cite{campostrini2006} yield very
similar results. In figure~\ref{fig:sfcdw}(b), we show the CDW structure
factor scaled in accordance with the 3D $XY$ universality class,
resulting in $V_c = 3.71(7)$.

We can check the robustness of this result by considering the condensate
fraction, which scales~\cite{binder1981,brezin1985,rath2010} according
to
\begin{equation}
  L^y f_c = g[(V-V_c) L^{1/\nu}],
\end{equation}
where $y = (d+z-2+\eta)$. This is illustrated in figure~\ref{fig:sfcdw}(c), 
resulting in $V_c =3.73(3)$. This result is quite close to the one obtained
using the structure factor. 
We stress, once again, that although this appears to be a second-order
transition between two ordered states, finite-size limitations do not
allow us to rule out the possibility of a weak first-order transition
or the existence of a small intermediate phase separating the BEC and
CDW states.

\begin{figure}[t]
  \centering
  \includegraphics[width=0.900\linewidth,draft=false]{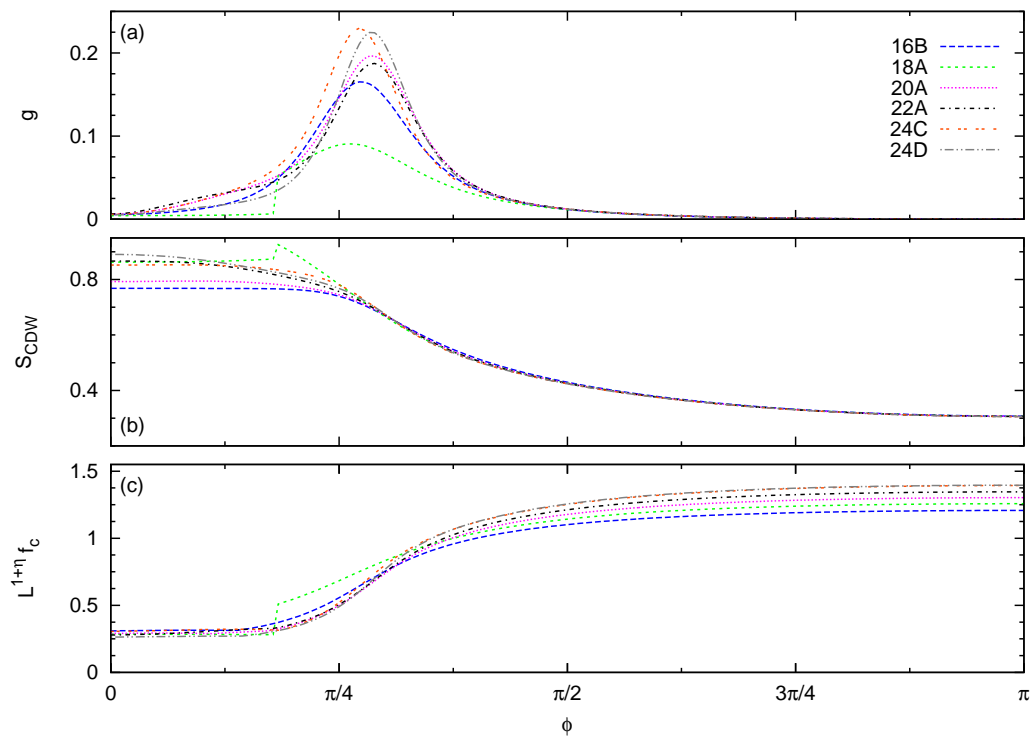}
  \caption{
    (Color online) (a) Fidelity metric $g$, (b) structure
    factor $S_{\rm CDW}$, and (c) scaled condensate fraction
    $L^{1+\eta} f_c$ as a function of $\phi$ for various clusters with
    $V = 0$. 
    \label{fig:sfbm}
  }
\end{figure}

Next, we examine the properties of the model as one transitions from
the Bose metal to the BEC state. In figure~\ref{fig:sfbm}, we show the
same quantities as in figure~\ref{fig:sfcdw} (this time versus $\phi$)
for $V=0$.  The fidelity metric is plotted in
figure~\ref{fig:sfbm}(a), and peaks at approximately $\phi\sim
0.88$. In figure~\ref{fig:sfbm}(b), we show the structure factor,
which does not scale with finite size in either phase. Figure
\ref{fig:sfbm}(c) depicts the scaled condensate fraction, yielding
$\phi_c = 0.84 \pm 0.14$, consistent with the peak in the fidelity
metric.  (Note that the $18A$ cluster experiences a level crossing for
$\phi<\phi_c$.)  As in the previous case, we cannot make definite
statements about the nature of the transition between the Bose metal
and the BEC state, but our results are consistent with a second order
or a weakly first order transition.

\begin{figure}[!t]
  \centering$
  \begin{array}{cccc}
    \includegraphics[width=0.230\textwidth,draft=false]{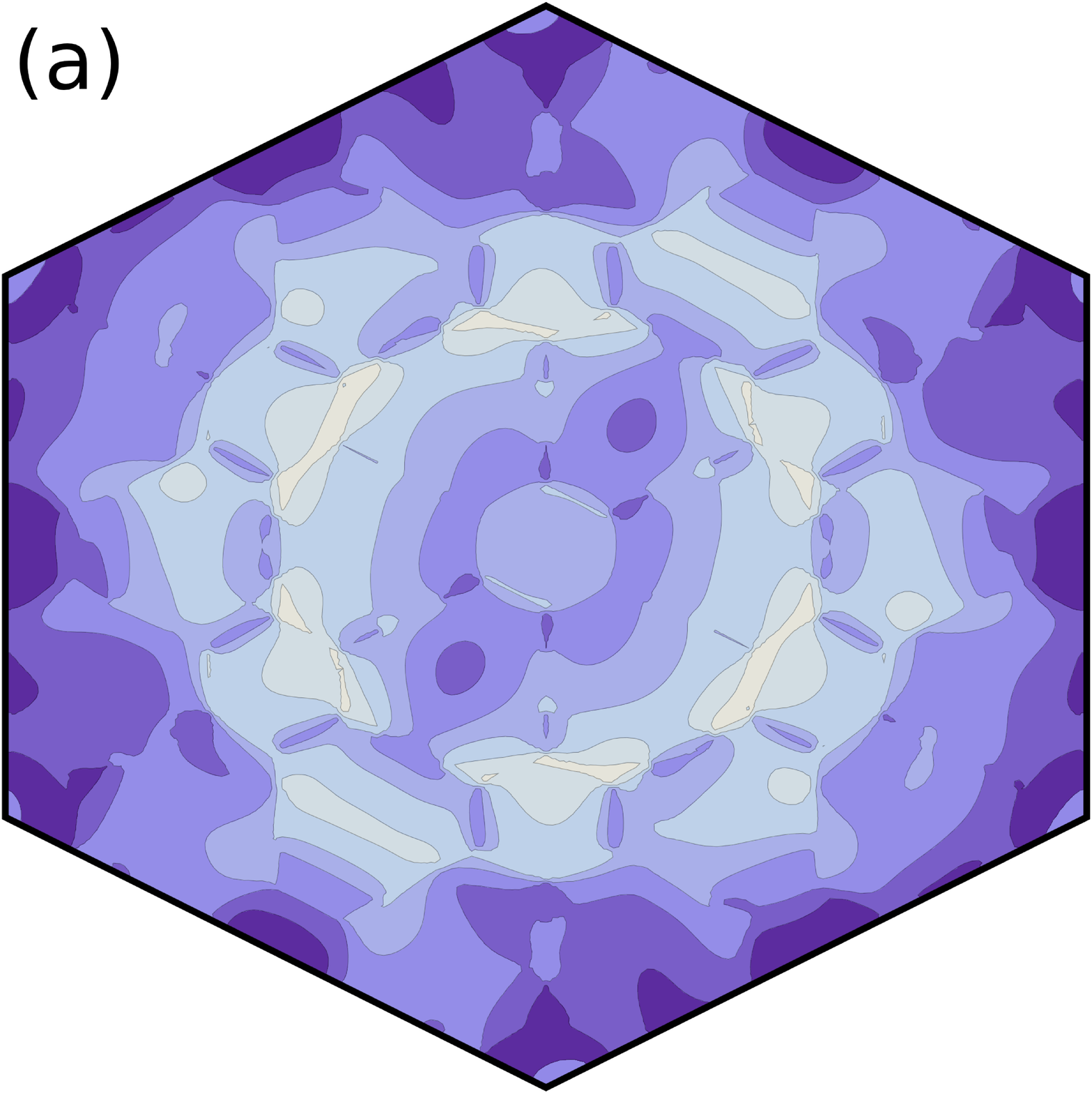} &
    \includegraphics[width=0.230\textwidth,draft=false]{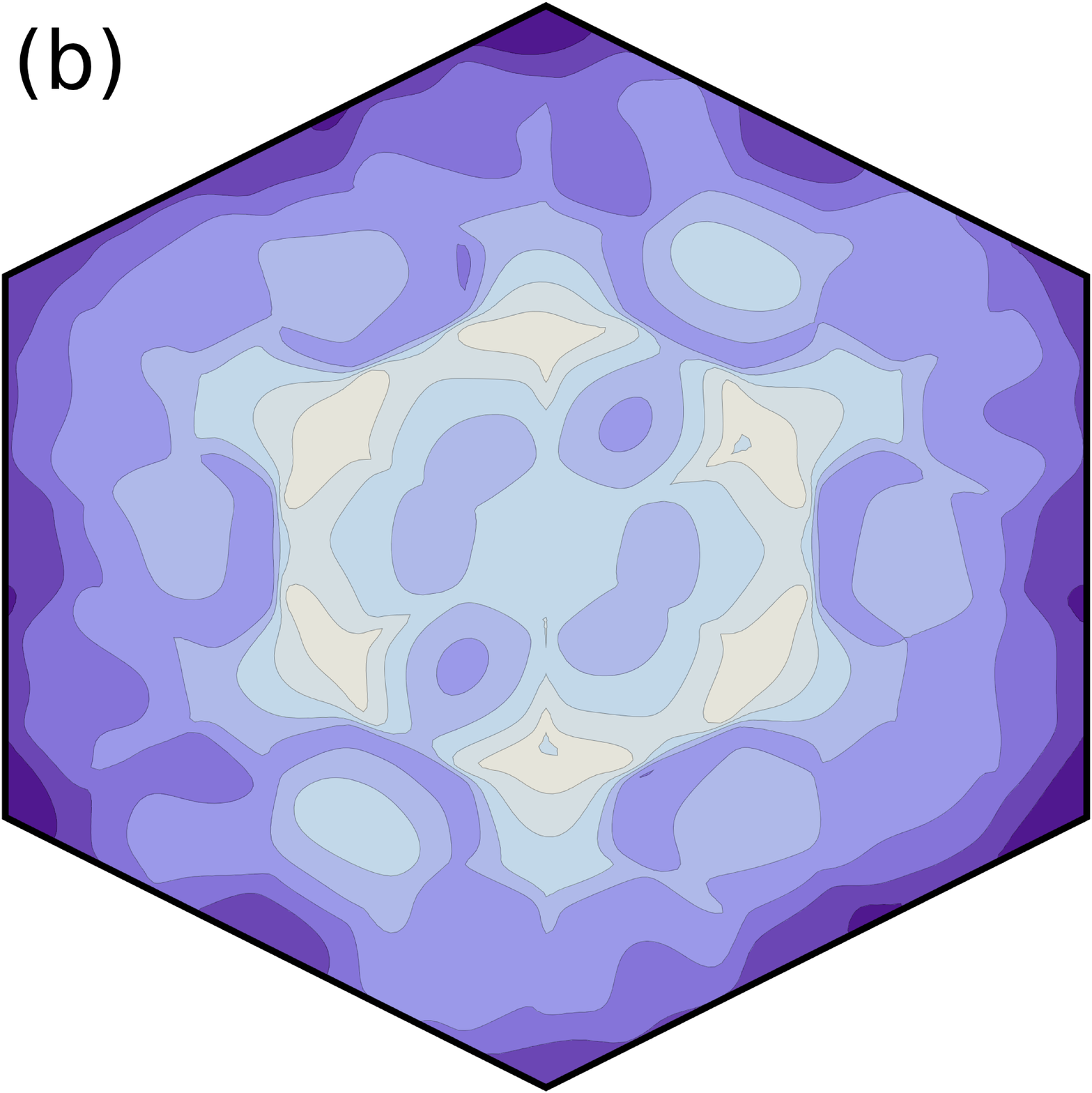} &
    \includegraphics[width=0.230\textwidth,draft=false]{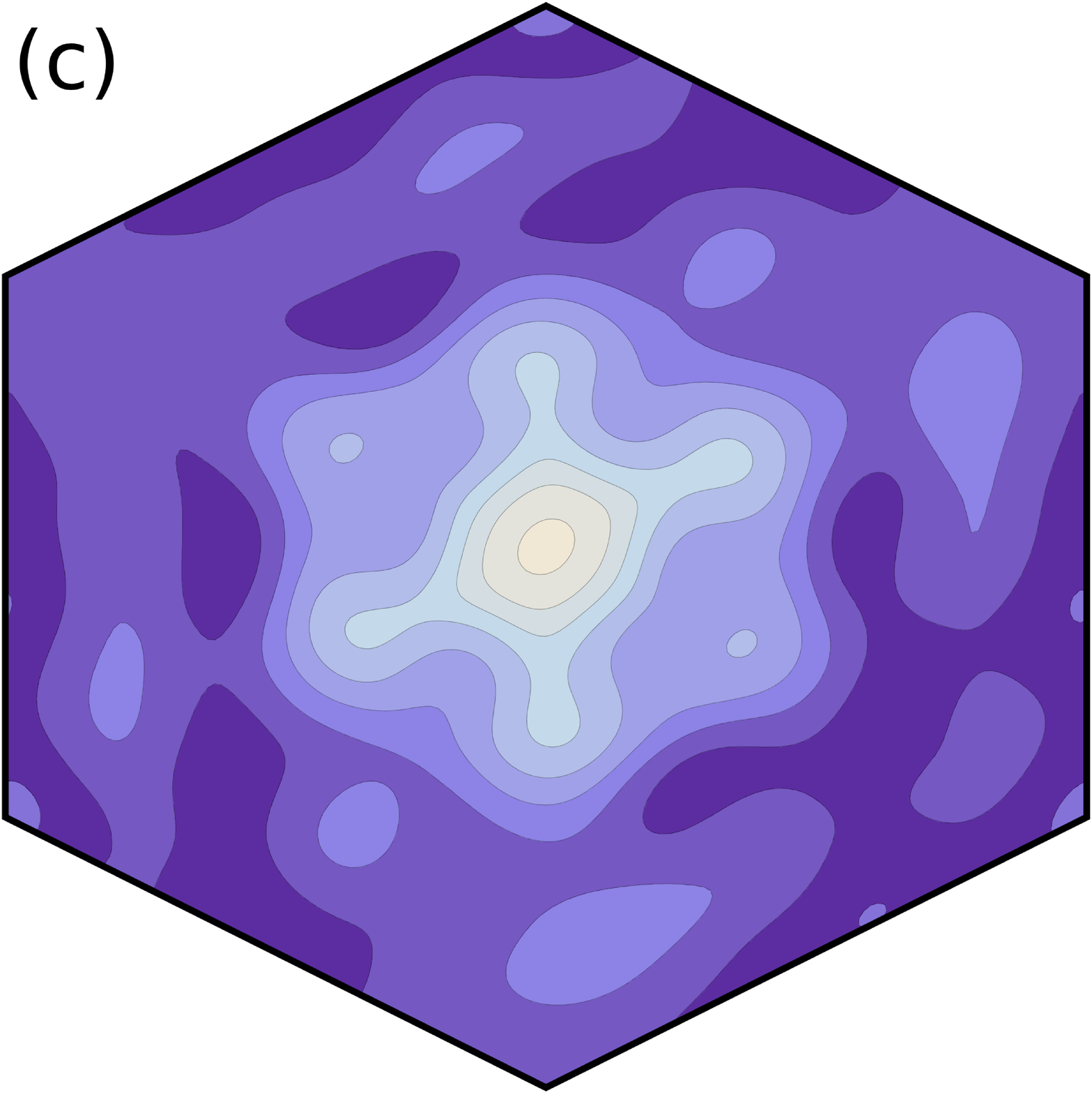} &
    \includegraphics[width=0.230\textwidth,draft=false]{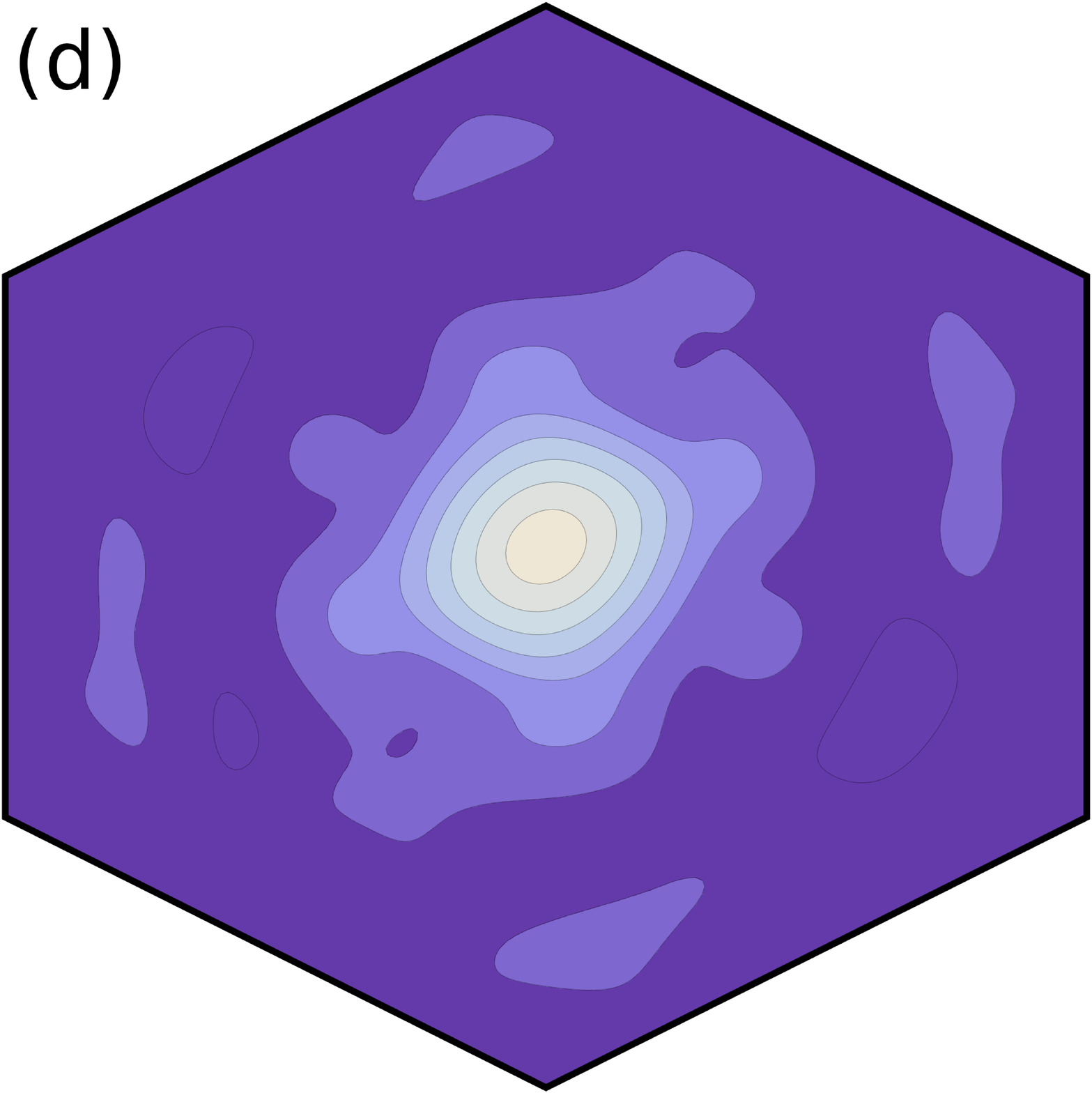}
  \end{array}$
  \caption{
    (Color online) Momentum distribution function $n({\bf k})$ versus
    ${\bf k}$~\cite{footnote2} for $V=0$ and (a) $\phi = 0$, (b) $\phi
    = \pi/9$, (c) $\phi = \pi/4$, and (d) $\phi = \pi / 3$. 
    \label{fig:nksfbm}
  }
\end{figure}

It is now interesting to study how the Bose surface changes as $\phi$
increases and one transitions between the Bose-metal and the BEC
phase. In figure~\ref{fig:nksfbm}, we show the momentum distribution
function for four values of $\phi$ and fixed $V=0$.  As seen in
figure~\ref{fig:nksfbm}(b), the Bose surface reduces in size as $\phi$
departs from zero and continues to shrink until condensation occurs at
${\bf k}=\Gamma$ [see Figs.~\ref{fig:nksfbm}(c) and
\ref{fig:nksfbm}(d)] for $\phi>\phi_c$.

\begin{figure}[!b]
  \centering
  \includegraphics[width=0.900\linewidth,draft=false]{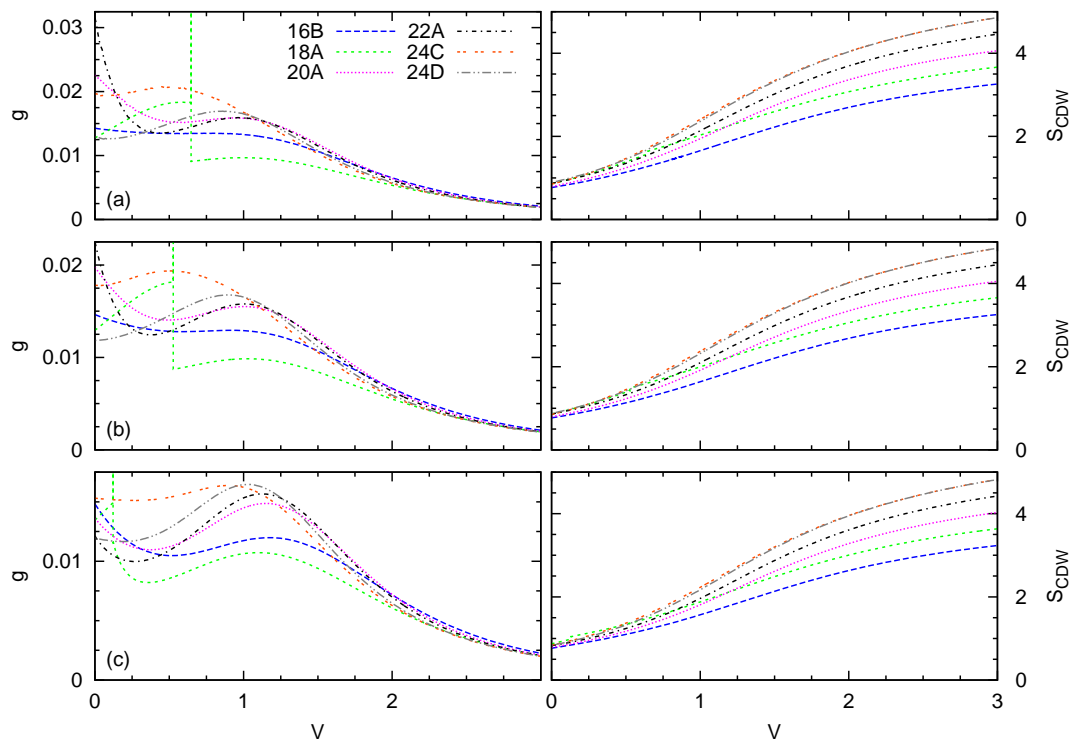}
  \caption{    
    (Color online) Fidelity metric $g$ (left) and structure factor
    $S_{\rm CDW}$ (right) as a function of interaction strength for
    various clusters with (a) $\phi = 0$, (b) $\phi = \pi / 12$, and
    (c) $\phi = \pi / 6$.
    \label{fig:bmcdw}
  }
\end{figure}

A third phase transition is expected as $V$ is increased from zero 
and the BM phase is destroyed to give rise to the large $V$ CDW phase.
We illustrate this regime in figure~\ref{fig:bmcdw} by plotting the fidelity 
metric (left panels) and CDW structure factor (right panels) for, 
(a) $\phi = 0$, (b) $\pi/12$, and (c) $\pi/6$. In the left panels, 
for all three values of $\phi$, one can see a sort of two-peak structure 
in the fidelity metric.

The large value of $g$ at $V = 0$ may be an indicator of a transition
away from the Bose metal at $V = 0^+$. It is somewhat similar to the
behavior of $g$ in both the one-dimensional Hubbard model, where the
Mott--metal-insulator phase transition occurs for the onsite repulsion
$U = 0^+$ \cite{campos2008}, and the two-dimensional hole-doped
$t$-$J$ model \cite{rigol2009}, where $d$-wave superconductivity was
seen to develop for a superconducting inducing perturbation with
vanishing strength. Another possibility is that the Bose-Metal is
stable for positive and small values of $V$, but a transition to
another phase occurs when $V<0$. The peak produced by such a
transition would also explain the structure we see in $g$. We have
also investigated this model with negative values of $V$, and found
that large peaks are present in the fidelity metric for $V<0$. The
position of those peaks had a strong dependence on the cluster
geometry. Hence, exactly what happens to the BM phase in the region
$V\sim 0$ is something that requires further studies, maybe with other
techniques that allow access to larger system sizes and a better
finite size scaling analysis.

For all clusters and values of $\phi$ depicted in the left panels 
in figure~\ref{fig:bmcdw}, one can also see a clear peak in the fidelity 
metric for finite values of $V$. This feature indicates the onset of 
CDW order. The structure factor $S_{\rm CDW}$, depicted in the right panels 
in figure~\ref{fig:bmcdw}, make apparent that for values of $V$ beyond 
that peak, the CDW structure factor scales with system size.

\begin{figure}[t]
  \centering$
  \begin{array}{ccc}
    \includegraphics[height=0.175\textheight,draft=false]{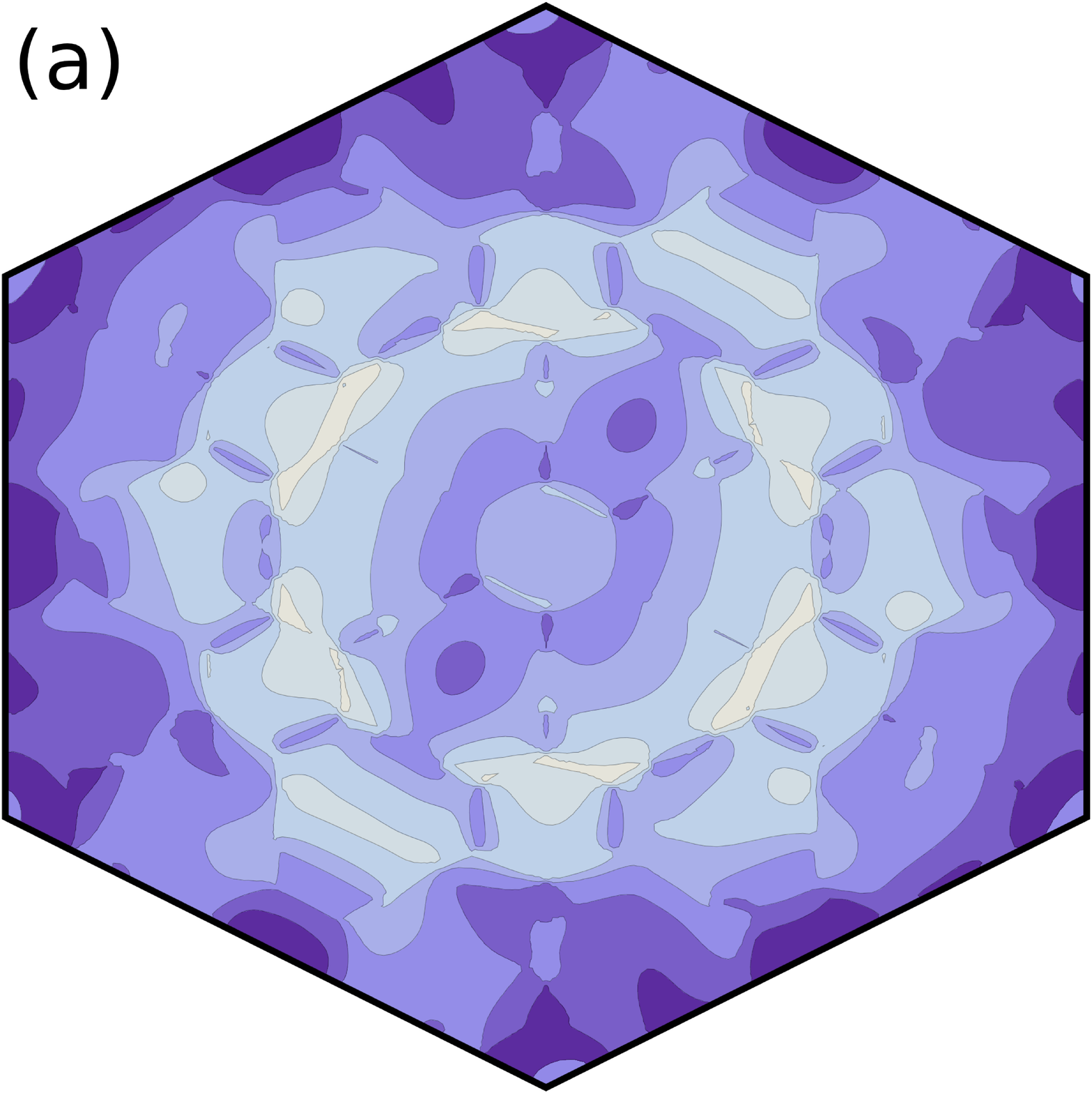} &
    \includegraphics[height=0.175\textheight,draft=false]{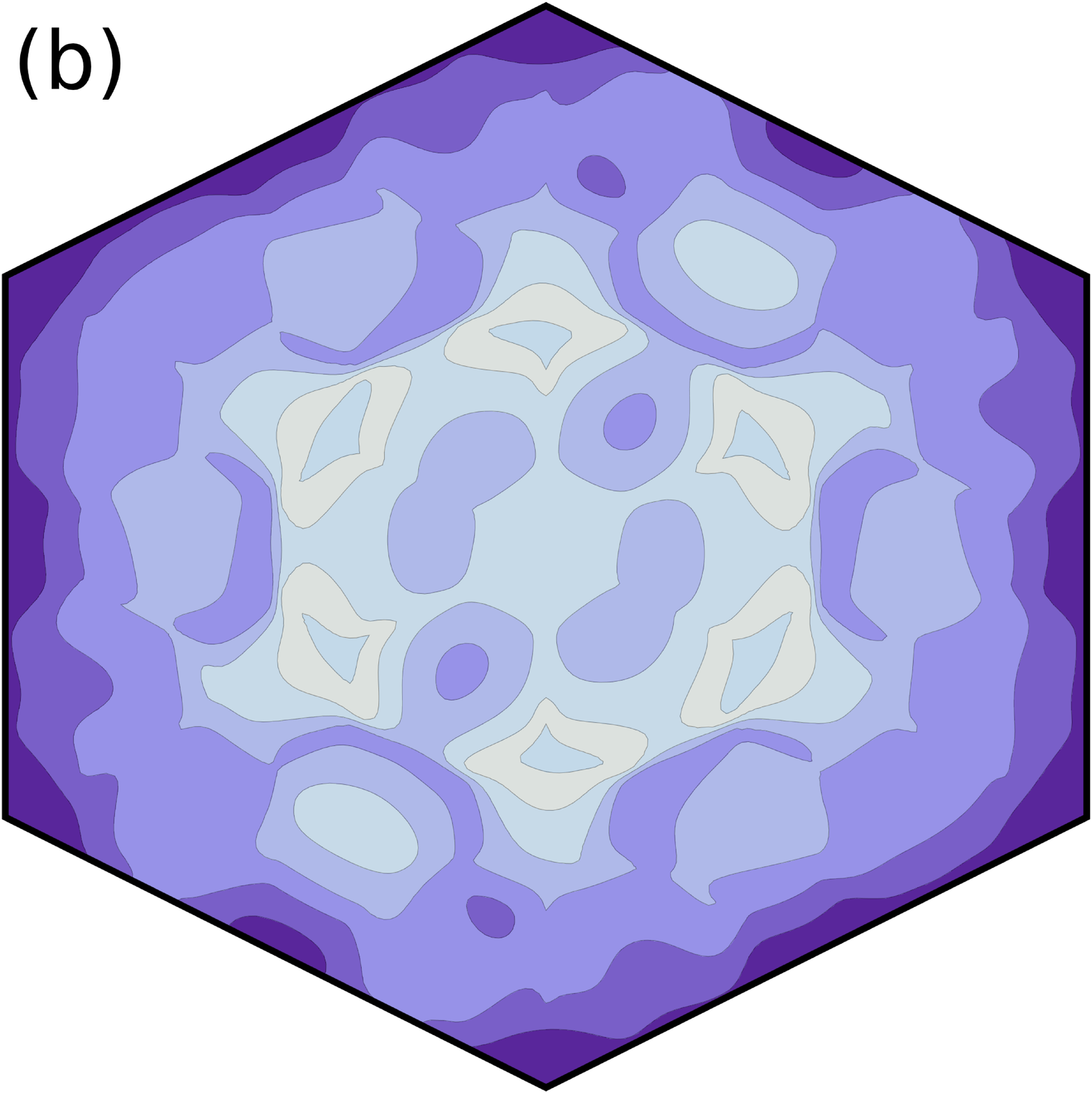} &
    \includegraphics[height=0.175\textheight,draft=false]{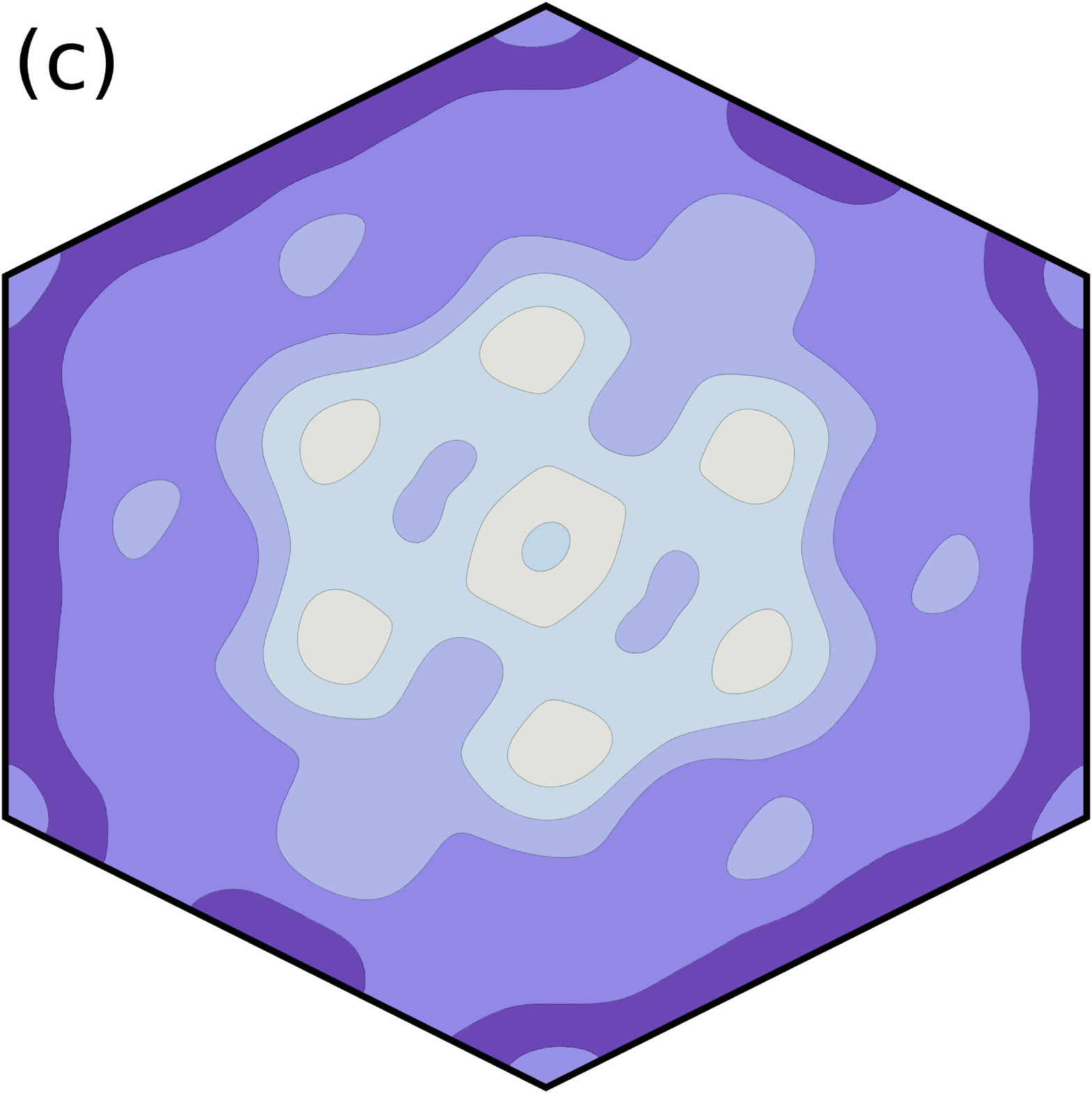} 
  \end{array}$
  \caption{
    (Color online) Momentum distribution function $n({\bf k})$ versus
    ${\bf k}$~\cite{footnote2} with constant $\phi = 0$ for (a) $V =
    0$, (b) $V = 0.25$ and (c) $V = 1$. 
    \label{fig:nkbmcdw}
  }
\end{figure}

In figure~\ref{fig:nkbmcdw}, we illustrate how the momentum distribution
function changes in the presence of interactions at fixed $\phi = 0$.
$n({\bf k})$ is shown for $V=0$ in panel (a) and as the interactions
are increased in panels (b) and (c). In figure~\ref{fig:nkbmcdw}(b), one
can see that the Bose-surface broadens as $V$ increases and moves
closer to ${\bf k} = \Gamma$. Increasing the nearest-neighbor
repulsion further, so that the system enters in the CDW phase
[figure~\ref{fig:nkbmcdw}(c)], results in a momentum distribution
function that peaked at ${\bf k}=\Gamma$, albeit without
condensation. Instead, the structure factor $S({\bf k})$ is sharply
peaked at ${\bf k} = \Gamma$.

\begin{figure}[t]
  \centering
  \includegraphics[width=0.700\linewidth,draft=false]{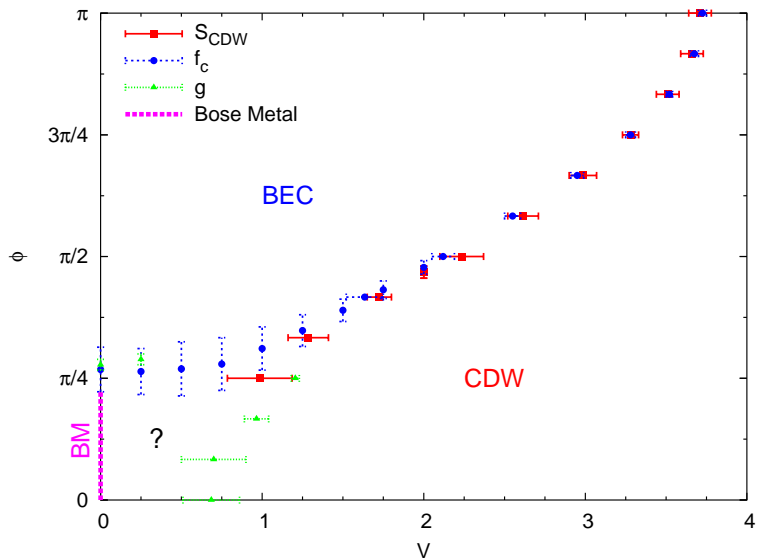}
  \caption{
    (Color online) Phase diagram for the Bose-Hubbard-Haldane model
    with parameters $t_1 = 1.0$ and $t_2 = 0.3$. The solid red squares
    are determined by the crossing point in the scaling of $S_{\rm
      CDW}$. The solid blue circles are from the crossing point in
    $f_c$. The green triangles are the average of the location of peak
    in the fidelity metric for the largest system sizes. 
    The Bose Metal (BM) phase is indicated by the
    thick, dashed magenta line. 
    \label{fig:bhhpd}
  }
\end{figure}

A summary of our calculations for different values of $V$ and $\phi$
is presented in figure~\ref{fig:bhhpd} as the phase diagram of the
hard-core BHH model at half-filling with $t_1=1.0$ and $t_2=0.3$.  For
$\phi > \pi / 4$, the boundary of the CDW phase was identified by the
crossing points in the scaling of the structure factor
(figure~\ref{fig:sfcdw}). The boundary of the BEC phase was identified
by the crossing points in the scaling of the condensate fraction
(Figs.~\ref{fig:sfcdw} and \ref{fig:sfbm}), and, for small values of
$V$, also using the maximum of the fidelity metric for the largest
systems sizes (figure~\ref{fig:sfbm}). For $\phi < \pi / 4$, the CDW
transition boundary was determined by the position of the maximum in
the peak in the fidelity metric for the largest system sizes
(figure~\ref{fig:bmcdw}). Note that, in that regime, the Bose metal
phase was found to be stable for $V = 0$. On the other hand, for $V$
between 0 and the boundary of the CDW phase, the large value of the
fidelity metric, as well as the behavior of several observables
studied in that region, prevent us from making a clear statement about
the nature of the ground state.

\section{Conclusion}
\label{sec:conc}
In summary, we have studied the phase diagram of the hard-core
Bose-Hubbard-Haldane Hamiltonian, which has allowed us to probe the
effect of perturbations on the Bose metal phase found in the
frustrated $XY$ model on a honeycomb lattice \cite{varney2011}. In
particular, we explored the parameter dependence of the Bose wave
vector and verified that the Bose metal is stable under the effects of
time-reversal and chiral symmetry breaking. We identified three phases
in the phase diagram of the BHH model; (I) a Bose metal, (II) a BEC,
and (III) a CDW. The phase transitions between the different phases
were identified utilizing the ground state fidelity metric, the CDW
structure factor, the condensate fraction, and the momentum
distribution.

The BEC-CDW transition appears to be second order, although
finite-size effects prevent us from ruling out the possibility of a
weak first-order transition or the existence of an intermediate phase
separating the BEC and the CDW states. If this transition is indeed a
direct second order phase transition, the critical point would be
highly nontrivial, and could be an example of deconfined criticality.

We have also found that the Bose metal is destroyed upon increasing
$V$, before the Heisenberg point for nearest neighbor interactions
$V=2J_1$ can be reached. The presence of a next-nearest-neighbor
repulsion may change this and result in transitions to other exotic
phases.

\ack
This research was supported by NSF through JQI-PFC (C.N.V. and K.S.),
ONR (C.N.V. and M.R.) and US-ARO (V.G.). The authors thank L. Balents,
M.~P.~A. Fisher, T.~C. Lang, and Z.~Y. Meng for useful discussions.

\bibliographystyle{iopart-num}
\bibliography{references}

\end{document}